\documentclass[12pt,titlepage,oneside]{revtex4}

\usepackage{amsmath}
\usepackage{amssymb}
\usepackage[dvips]{graphicx}
\usepackage{psfrag}
\usepackage{pstricks}
\usepackage{pst-node}
\usepackage{pst-plot}

 \psfrag{(a)}{(a)}
 \psfrag{(b)}{(b)}
 \psfrag{(c)}{(c)}
 \psfrag{(d)}{(d)}
 \psfrag{X}{$X$}
 \psfrag{Z}{$Z$}
 \psfrag{1}{$1$}
 \psfrag{Ze1}{$Z=1$}

\newcommand{\qed}{\hspace*{\fill}$\square$}

 \newcommand{\Z}{\mathbf{Z}}

 \newcommand{\sset}[1]{ \{#1\} }

 \newcommand{\prima}{^\prime}
 
 \newcommand{\daga}{^\dagger}

 \newcommand{\ket}[1]{|#1\rangle}

 \newcommand{\simS}{\sim_{\mathcal S}}
\newcommand{\simH}{\sim_{H}}

\begin{document}

\title[Short Title]{
Quantum Measurements and Gates by Code Deformation}

\author{H. Bombin and M.A. Martin-Delgado}
\affiliation{
Departamento de F\'{\i}sica Te\'orica I, Universidad Complutense,
28040. Madrid, Spain.
}

\begin{abstract}
The usual scenario in fault tolerant quantum computation involves
certain amount of qubits encoded in each code block, transversal
operations between them and destructive measurements of ancillary
code blocks. We propose to complement this techniques with code
deformation, in which a given code is progressively changed in such
a way that encoded qubits can be created, manipulated and
non-destructively measured. We apply this approach to surface codes,
where the computation is performed in a single code layer which is
deformed using `cut and paste' operations. All the interactions
between qubits remain purely local in a two-dimensional setting.

\end{abstract}


\maketitle

\section{Introduction}

Quantum computing is the art of controlling quantum coherence at
will. It is no surprise then that the main obstacle towards this
enormous achievement is the decoherence that any real-life quantum
system will suffer unavoidably. The quest to overcome this
difficulty, first believed to be insurmountable, led to the creation
of quantum error correcting codes \cite{shor95}, \cite{steane96a}.
These codes revolve around the idea of storing a small amount of
quantum information in a large number of quantum degrees of freedom,
so that the redundancy makes the information more resilient to
errors. Among the most fruitful ones are stabilizer quantum error
correcting codes \cite{gottesman96}, \cite{calderbank_etal97}, which
are particularly easy to analyze. Some of them even allow to
implement interesting sets of gates transversally, that is, in such
a way that localized errors do not spread uncontrolledly over the
code block.

Transversal gates have proved themselves very useful, but sometimes
they are not enough. In this paper we consider the concept of code
deformation, which enlarges the amount of fault tolerant gates
naturally implementable in a given code. The idea is that a quantum
error correcting code can be slightly changed to give a new one, and
that such changes can be applied one after the other to produce
quantum gates on encoded qubits. The amount of encoded qubits can
change due to deformations, giving rise to initialization and
measurement operations. As we will see such ideas are natural in the
context of topological codes, but they could be applied in other
kinds of quantum error correcting codes.

An important practical problem regarding many theoretical proposals
involving quantum error correction and fault tolerant computation is
the non-locality of the elementary gates between the physical qubits
that make up the codes. In some practical devices, elementary gates
have a local nature, a severe restriction when errors are to be
taken into account. Here dimensionality is another important issue,
since it is not the same thing to have local gates in a 1D, 2D or 3D
system.

Surface codes, introduced by Kitaev \cite{kitaev97}, are stabilizer
codes with the interesting feature of being local in a 2D setup. In
particular, the qubits can be arranged in a 2D lattice in such a way
that the necessary measurements to control the errors only involve
as few as four neighboring qubits. Then one can envision a quantum
computer as a stack of layers. Each layer is a surface code encoding
a single qubit, and CNot gates are performed transversally in pairs
of layers. The problem with surface codes is that no other gate can
be transversally implemented. To overcome this difficulty color
codes were developed \cite{topologicalclifford}, which are also
local in 2D, but allow the transversal implementation of the whole
Clifford group, which is enough for quantum distillation and many
other quantum information tasks, such as quantum teleportation.
Extensions of color codes to 3D with nice transversal properties for
universal quantum computation are also possible \cite{tetraUQC}.

However, working with a 2D setup is a rather typical scenario in
technological applications such as lithographic techniques and the
like. In that case, the above setting with several stacks of planar
codes becomes useless. Then one can wonder wether it is possible to
implement non-trivial gates by means of some different scheme. In
this paper we answer this question in the positive. In particular,
we show how non-destructive measurements and CNot gates can be
obtained by deforming a single-layered surface code.
This is in sharp contrast with the standard approach to surface
codes.

The idea of inserting and removing physical qubits from a surface
code was introduced in \cite{dennis_etal02} as a way to correct
transversal Hadamard gates. This insertion an removal gives rise to
a progressive reshaping of the lattice or, what is the same, a
deformation of the code. We want to stress that indeed there is a
fixed underlying lattice of qubits. When we say that a site is added
to or removed from the surface code, we do not mean a change in the
underlying physical system but just on the stabilizers that define
our code. In a sense, what we are reshaping is a `software' lattice,
whereas the underlying `hardware' lattice remains intact.

Similar ideas have already been discussed in the context of cluster
states \cite{rauss}. Our approach differs in several aspects. First,
our discussion is based solely on the properties of 2D topological
codes with no additional structure, which makes it clearer. It is
natural for us to consider not only surfaces with holes, but really
any kind of topologies, including those in which two types of
boundaries appear next to each other. The way in which we propose to
perform deformations fits naturally in the error correction scheme
discussed in \cite{dennis_etal02}, and we show how the gates
discussed in that work can be applied in our context. Moreover, we
discuss code deformation in such a way that it can be inmediately
applied to other codes, for example those in
\cite{topologicalclifford}, \cite{tetraUQC}.

Regarding approaches to surface codes for topological quantum computation
processing, we want to stress that throughout this paper we are considering
the scenario of active quantum error correction introduced in
\cite{shor95,steane96a} but applied to codes based on topological properties
of a quantum system. The benefits that we want to explore in this approach
come from the nice locality properties of surface codes and the metods for
code deformation that we introduce.
This is sometimes referred as quantum protection at the software level.
Yet, there is another approach to
topological quantum computation also envisioned by Kitaev \cite{kitaev97}
that considers the possibility of achieving the protection of quantum
states at the hardware level. The underlying idea in this alternative proposal
is that the quantum topological properties of the system may yield
a self-protecting scenario in which, when an error pops up in the code,
then the system is so well-protected topologically that it can correct
those errors internally by itself, without resorting to external means
as in the active error correction scenario. The initial reasons to
believe that surface codes could lead to this type of self-protecting
robustness were based on both the resistance to local errors versus
global encoding along with the existence of a protecting energy gap.
This gap is allegedly the physical mechanism for the topological protection
since the quantum code is represented by the degenerate ground state of a
quantum lattice Hamiltonian and the excited states amount to errors in
the quantum topological code. However, Kitaev Hamiltonians are not
stable with respect to thermal noise. This fact was already noticed
in \cite{dennis_etal02}, and more recently, a final rigorous proof
has been given in \cite{AFH08}.
The only known model which shows thermal stability is the
4D Kitaev model \cite{AHHH08}.

In summary, we want to stress that in this paper our approach is that
of active quantum error correction. In particular, this means that
we are not considering any Kitaev Hamiltonian associated to our
quantum codes. Instead, we perform external operations to achieve
quantum error corrections on systems where the encoding of information
is done at the topological level.

\section {Code deformation}

An error correcting code is a subspace of the space of states of a
given quantum system. The error correcting capabilities are related
to the fact that certain operators or errors, those that the code
can detect, do not connect orthogonal encoded states. We want to
consider small changes that progressively deform one stabilizer code
into another. The motivation is that such deformations can be used
to initialize, transform and measure encoded qubits.

\subsection {Stabilizer codes}

Given a certain number of qubits, its group of Pauli operators
$\mathcal P$ is defined as that generated by tensor products of the
usual $X$ and $Z$ single-qubit Pauli operators. For example, if we
have five qubits, a generator would be $X\otimes 1 \otimes Y\otimes
Z\otimes X$. A stabilizer code \cite{gottesman96},
\cite{calderbank_etal97} of length $n$ is a subspace of the quantum
system of $n$ qubits. It is described as the subspace $\mathcal C$
stabilized by an Abelian subgroup $\mathcal S\subset \mathcal P$.
The stabilizer group $\mathcal S$ must not contain $-1$ as an
element, and if it has $n-k$ independent generators $S_i$, the
encoded subspace $\mathcal C$ has dimension $2^k$ and thus we say
that it encodes $k$ qubits. The encoded states $\ket\psi\in \mathcal
C$ are characterized by the conditions $S_i\ket\psi=\ket\psi$,
$i=1,\dots, n-k$.

An important tool in the understanding of stabilizer codes is the
normalizer $\mathcal N$ of $\mathcal S$. This is the subgroup of
Pauli operators that commute with all the elements of $\mathcal S$.
Define the weight of a Pauli operator as the number of non-trivial
terms in its tensor product expression. Then, the minimum of the
weights of the elements of $\mathcal N-\mathcal S$ gives the so
called distance of the code. This is the minimun number of qubits
that we have to manipulate in order to change one encoded state into
another. Thus, the bigger this distance, the bigger the noise
resilience of the code. From the normalizer one can always choose
elements $X_i$, $Z_i$, $i=1,\dots, k$ such that
\begin{equation}\label{encoded_paulis}
[X_i, X_j]=0, \qquad [Z_i, Z_j]=0,\qquad X_i Z_j=(-1)^{\delta_{i,j}} X_j Z_i.
\end{equation}
These generate the group of encoded Pauli operators, that is, the Pauli
operators of the encoded qubits.

\subsection{Code transformations}

Consider two codes $\mathcal C$, $\mathcal C\prima$ with stabilizers
$\mathcal S$, $\mathcal S\prima$, both with the same number of
physical qubits $n$ and encoded qubits $k$. Let us denote the
generators of the stabilizers as $S_i$, $S_i\prima$, and the
elements of the basis of encoded Pauli operators as $X_i, Z_i$ and
$X_i\prima, Z_i\prima$, respectively. The Clifford group
\cite{nielsen_chuang} consists of those unitary operators $U$ in our
system of $n$ qubits such that for any element $T\in \mathcal P$ we
have $UTU\daga\in \mathcal P$. Consider the subset of Clifford
operators $U$ with $U\mathcal S U\daga= \mathcal S\prima$, that is,
those that transform $\mathcal C$ into $\mathcal C\prima$. Such
operators are very general. For example, they include transversal
operations \cite{shor96}, \cite{gottesman96}, in which $\mathcal
C\prima = \mathcal C$ and $\mathcal U=u^{\otimes n}$ with $u$ some
unitary single-qubit operator. Transversal operations have a great
importance in fault-tolerant quantum computation \cite{knill98},
\cite{zeng}, but it is interesting to look for alternatives that can
widen the applicability of fault-tolerant codes. Here we explore the
idea of code deformations, in which only some of the generators of
the stabilizer change. Then, if $r$ of them change we can write
$S_i\prima=US_i U\daga  = S_i$ for $i=r+1,\dots, n-k$. If this $r$
is somehow small, it makes sense to talk about code deformations. We
will see how in surface codes deformations have indeed a geometrical
meaning, because they take the form of localized changes in the
shape of a given surface. Because these changes do not alter the
topology of the surfaces, we call them smooth deformations.

When $\mathcal C=\mathcal C\prima$, we have
\begin{equation}\label{evolucion_paulis}
U X_i U\daga  \simS \prod_j X_j^{a_{ij}} Z_j^{b_{ij}},\qquad U Z_i
U\daga  \simS \prod_j X_j^{c_{ij}} Z_j^{d_{ij}},
\end{equation}
where $a_{ij},b_{ij},c_{ij},d_{ij}\in \Z_2$ and $A\simS B$ if $A=BS$
for some $S\in \mathcal S$. The evolution of the encoded Pauli
operators \eqref{evolucion_paulis} determines, up to a phase, the
unitary evolution of the encoded qubits under $U$. In the case of
code deformations, which change the code only partially, we can
successively apply several operators $U_i$ so that $U=U_t\dots U_1$
takes the code into itself. Thus, we can use deformations to perform
Clifford gates, as we will see in particular in surface codes.

We have said that both codes, the initial and the final, have the
same number of physical qubits. However, it should be noted that we
can use the previous description also in a case in which the numbers
differ. This is so because we can always enlarge any code with
additional qubits. We simply add one stabilizer generator per new
qubit, each one, for example, a $Z$ on the corresponding qubit. This
way, extra qubits are in a fixed state for encoded states and thus
do not affect the code.

Indeed, it was the need to enlarge an existing code qubit by qubit
which motivated the introduction of code deformations in
\cite{dennis_etal02} for surface codes. In the mentioned work the
operator $U$, which amounts to several local CNot gates, is applied
explicitly to the code in order to deform it. Alternatively, one can
perform the deformation by measuring the new stabilizers
$S_i\prima$, $i=1,\dots,r$. Some of the measurements can give an
undesired value, so that the obtained code has stabilizers $m_i S_i$
with $m_i=\pm 1$. This could then be corrected applying a suitable
Pauli operator, as we will see in specific examples. In practice,
however, this correction is unnecessary, at least when error
correction reduces to monitor errors in the system, as in
\cite{dennis_etal02}. In that case, one performs round after round
of measurements. In each round, all the generators $S_i$ are
measured. These measurements are used to calculate with high
confidence which errors occurred. When some measurement is done on
encoded qubits, it must be interpreted taking these errors into
account, which are never really corrected in any other way.
Performing deformations by measuring the stabilizers fits nicely in
this error correction scheme, because we can change the stabilizers
to be measured from one round to another in order to get the desired
deformations. However, for this deformation procedure to work it
must be possible to correct errors successfully with high
probability. In other case, one has to apply directly a suitable
unitary operator. In the case of surface codes, deformations can be
done through measurements as long as we keep them local when
compared to the support of encoded Pauli operators, except for
encoded operators which are being initialized or measured, as we
discuss in the next section.

\subsection{Initialization and measurement}

The code transformations discussed in the previous section cannot
change the number of encoded qubits. They can be done without
introducing stabilizer violations, simply by implementing $U$
suitably. But we can consider more general Clifford operators $U$,
in particular such that the initial code $\mathcal C$ and the final
code $\mathcal C\prima$ have a different number of encoded qubits.
The condition $U\mathcal S U\daga =\mathcal S\prima$ can no longer
be imposed. In the case of deformations in surface codes, as will
see, these correspond to changes in the topology of the surface, and
thus we call them non-smooth deformations.

Suppose first that $\mathcal C$ encodes $k$ qubits and $\mathcal
C\prima$ encodes $k+1$ qubits. Then $\mathcal C\prima$ has one
stabilizer generator less. We consider those Clifford operators $U$
with $\mathcal S\prima\subsetneq U\daga \mathcal S U$, that is,
those which transform encoded states into encoded states. Then
$U\mathcal S U\daga $ is a subset of $\mathcal N\prima$ generated by
$\sset{S_i}_{i=1}^{n-k}\cup\sset{T}$ where $T$ is a nontrivial
encoded Pauli operator, $T\in \mathcal N\prima- \mathcal S\prima$.
This $T$ has eigenvalue one after $U$ has been applied. Thus, $U$
not only adds one qubit but also initializes it in a definite way.
We can consider in a similar way the introduction of several new
qubits and their initialization. In the particular case of surface
code deformations, such operators $U$ will introduce changes in the
topology which increase the number of nontrivial cycles.

Now consider the reverse case, so that $\mathcal C$ encodes $k$
qubits and $\mathcal C\prima$ encodes $k-1$ qubits. Then $\mathcal
C\prima$ has one stabilizer generator more. We consider those
Clifford operators $U$ with $U\mathcal SU\daga\subsetneq \mathcal
S\prima $, that is, those which are inverses of the transformations
just considered. Then, $U\daga\mathcal S\prima U$ is a subset of
$\mathcal N$ generated by $\sset{S_i}_{i=1}^{n-k}\cup\sset{T}$ where
$T$ is a nontrivial encoded Pauli operator, $T\in \mathcal N-
\mathcal S$. This $T$ is mapped onto an element of $\mathcal
S\prima$ through $U$, and in this sense gets measured. Thus, $U$
removes one qubit and the corresponding degrees of freedom map onto
possible violations of the stabilizer in the new code. Again, we can
consider the removal and measurement of several qubits at the same
time. For surface codes, such operators $U$ will introduce changes
in the topology which decrease the number of nontrivial cycles.

It should be noted that after initializing $T$ or before measuring
it we do not care if the code suffers a $T$ error. In the first case
because the encoded state satisfies $T=1$, so that the action of $T$
is trivial. In the second case, because the error will create
decoherence between states with $T=1$ and $T=-1$, which is
irrelevant since we intend to measure $T$. This discussion is
pertinent for surface codes, since changes in the topology will
expose the code to errors, but these errors will always correspond
to the operators being measured or initialized. In a surface code, the
protection of quantum information is obtained by storing it in
non-local degrees of freedom. Thus, it is unavoidable that when we
perform an abrupt change in the topology of the surface some degrees
of freedom become local. Fortunately, this causes no harm, because to
initialize or measure an encoded operator $T$ we only make $T$
local.

\section{Surface codes}

\begin{figure}
 \psfrag{a}{a}
 \psfrag{b}{b}
 \psfrag{c}{c}
 \psfrag{d}{d}
 \psfrag{e}{e}
 \psfrag{f}{f}
 \psfrag{g}{g}
 \includegraphics[width=14cm]{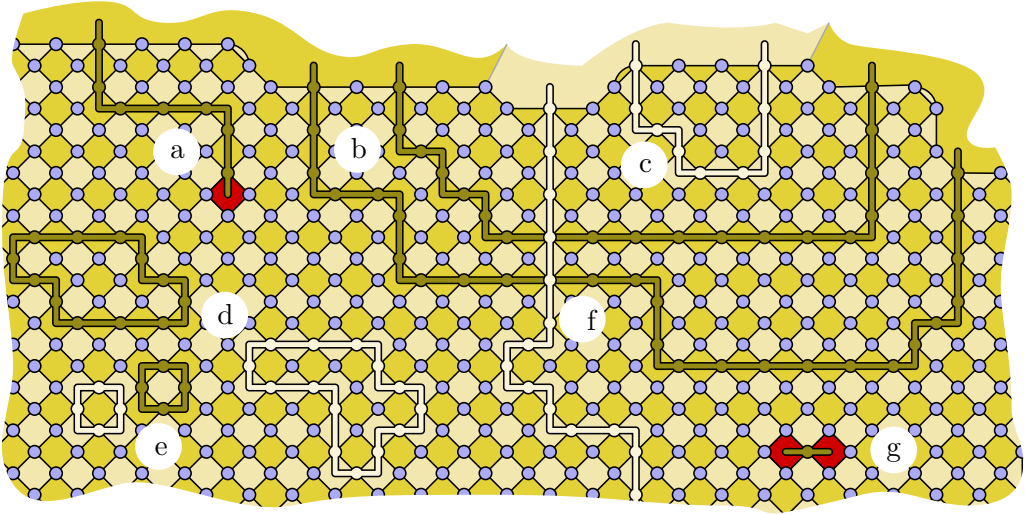}
 \caption
 {
A piece of a surface code with an irregular dark-light-dark border
on the top part. There is a qubit at each site of the lattice. Dark
(light) strings represent products of $Z$ ($X$) operators. (e) Plaquette operators as strings. (c,
d) Products of plaquette operators are boundary strings. (a, g) If a
string operator has an endpoint at a plaquette, it will not commute
with the plaquette operator. (b) If two strings operators are equal
up to deformation, their action on encoded states is the same. (f)
Crossing $X$ and $Z$ string operators anticommute.
 }
 \label{figura_surface_codes}
\end{figure}

In order to construct surface codes one starts considering any
two-dimensional lattice in which four links meet at each site and
plaquettes can be two-colored. For our purposes and to fix ideas, a
`chessboard' lattice in the plane will mostly suffice, see
Fig.~\ref{figura_surface_codes}. This lattice will have borders,
which can be best understood as big missing faces. Since there are
two kinds of plaquettes, dark and light, borders have also this
labeling. In the interface between a dark and a light border there
are missing edges that would separate the missing plaquettes.

To construct the quantum error correcting code from the lattice, a
physical qubit must be placed at each site. The stabilizer $\mathcal S$ is
generated by plaquette operators. Given a plaquette $p$ we define
the plaquette operator $X_p$ ($Z_p$) as the tensor product of
$X$ ($Z$) Pauli operators acting on those qubits lying on the
plaquette. Then we attach $X_p$ operators to dark plaquettes and
$Z_p$ operators to light plaquettes. All these operators commute
due to the properties of the lattice, and thus the stabilizer is well defined.

Any operator which is just a tensor product of $Z$ ($X$) operators
acting on certain qubits can be visualized as a string that goes
through the corresponding sites and lives on dark (light)
plaquettes, see Fig.~\ref{figura_surface_codes}. So given a dark
(light) strings $s$, i.e. a geometric object, we attach to it the
operator $Z_s$ ($X_s$). If a light (dark) plaquette $p$ is
considered as a small closed dark (light) string, this definition
agrees with the previous one, see Fig.~\ref{figura_surface_codes}
(e). For a closed string we mean a string with no endpoints. Given a
dark (light) string operator $s$ and a dark (light) plaquette $p$,
$[Z_s,X_p]=0$ ($[X_s,Z_p]=0$) iff $p$ is not an endpoint of $s$, see
Fig.~\ref{figura_surface_codes}(a,g). Thus, if $s$ is dark (light)
closed string then $Z_s\in \mathcal N$ ($X_s\in \mathcal N$). As
borders are indeed missing plaquettes, a dark (light) string $s$
with its endpoints at dark (light) borders should be considered a
closed string because $Z_s\in \mathcal N$ ($X_s\in \mathcal N$). In
terms of homology, the homology of dark (light) strings is a
homology relative to dark (light) borders
\cite{bravyi_kitaev_bordes}, \cite{bmd_homologia}.

The previous observations imply that closed string operators
generate the normalizer $\mathcal N$ of the stabilizer $\mathcal S$.
Moreover, $\mathcal S$ is generated by boundary string operators. A
dark (light) boundary string is a string which encloses some portion
of the surface, which can include dark (light) borders but not light
(dark) ones. Then the elements of $\mathcal N-\mathcal S$ take the
form $X_sZ_{s\prima}$, with $s$, $s\prima$ closed and at least one
of them not a boundary. The elements of the basis of encoded Pauli
operators, $X_i, Z_i\in \mathcal N-\mathcal S$, can be chosen
graphically. To this end, two points must be taken into account.
First, when two dark (light) strings $s$, $s\prima$ of the same type
differ only by a boundary, see Fig.~\ref{figura_surface_codes}(b),
the corresponding operators are equivalent up to products with
stabilizer  elements, $Z_s \simS Z_{s\prima}$ ($X_s \simS
X_{s\prima}$). When we say that $s$ and $s\prima$ differ only by a
boundary we mean that they can be deformed one into the other or,
more exactly, that they are equivalent up to $\Z_2$ homology,
$s\simH s\prima$. Secondly, if $s$ is a dark string and $s\prima$ a
light string, $\sset{X_s,Z_s}=0$ iff $s$ and $s\prima$ have an odd
number of sites in common or, what is the same, they cross an odd
number of times, see Fig.~\ref{figura_surface_codes}(f).

\begin{figure}
 \includegraphics[width=10cm]{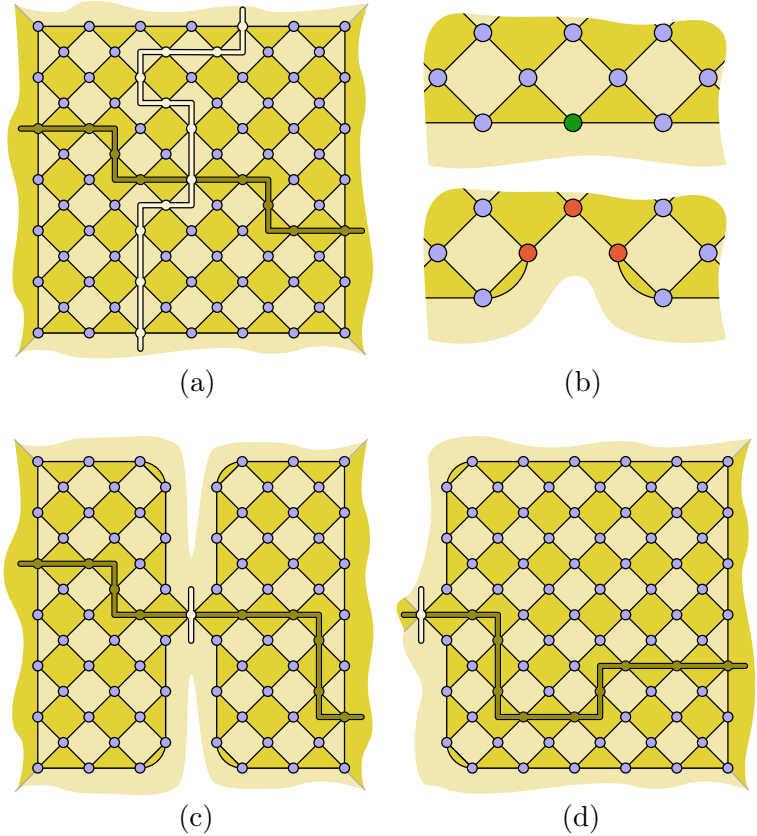}
 \caption{
 (a) A surface code with a dark-light-dark-light
 border structure. It encodes a single qubit. Nontrivial strings
 that correspond to the encoded
 $X$ and $Z$
 operators are displayed.
 (b) An elementary deformation.
 In case that after the removal of the green qubit
 the new two-sited plaquette operators have negative eigenvalue, $Z$
 operators are applied at red qubits.
 (c) The previous code after being deformed so that
 the encoded $X$ becomes exposed to local measurements.
 (d) A dark border was reduced to expose the encoded $X$.
 }
 \label{figura_codigo_simple}
\end{figure}

All this is best illustrated with an example. Consider the surface
code depicted in Fig.~\ref{figura_codigo_simple}(a). It encodes a
single qubit. Let $s$ be the dark string and $s\prima$ the light
one. These strings are closed but not boundaries, and any other
homologycally non-trivial string is equivalent to one of them. The
Pauli operator basis is given by $Z_1=Z_s$ and $X_1=X_{s\prima}$.
The bigger the width and height of the rectangular surface are (in
terms of qubits), the more resilient the code will be to $Z$ and $X$
errors, respectively.

\section{Surface code deformation}

We are now in position to discuss deformations in surface codes. As
we have already mentioned, these take the form of geometrical
changes in the shape of the surface. As qubits enter and exit the
code, the dark and light borders of the surface change. We can move
the borders, glue them together or create new ones cutting the
surface. Although we talk about introducing and removing qubits from
the surface, there is a fixed underlying square lattice of physical
qubits. We change the stabilizers that define the code, not the
physical system.

The basic mechanism is exemplified in
Fig.~\ref{figura_codigo_simple}(b), where a qubit in a light border
is erased, causing the removal of a light plaquette operator and the
change of two dark plaquette operators. To perform the deformation
of the code, the new two-sited plaquette operators must be measured.
In the absence of errors, their eigenvalues must agree. If they are
negative, we can apply $Z$ operators on those qubits marked in red.
In practice, errors may appear and we just perform the measurements,
which will then be interpreted at the error correction stage
\cite{dennis_etal02}. If the inverse deformation of that of
Fig.~\ref{figura_codigo_simple}(b) is performed, corrections have to
be made on the qubit marked in green.

\subsection {Smooth deformations}

First we want to consider smooth deformations, in which the topology
of the surface is not altered. Such transformations cannot change
the number of encoded qubits, but can perform unitary gates on them.
So suppose that we deform a surface in an arbitrary way, taking it
finally back to its original form. The total deformation gives a
continuous mapping $f$ of the original surface onto itself. In
particular, this mapping takes strings $s$ to $f(s)$. Recall that
encoded Pauli operators can be chosen to be string operators.
Moreover, we can find light strings $s_i$ and dark strings
$s_i\prima$ so that $X_i:=X_{s_i}$ and $Z_i:=Z_{s\prima_i}$ satisfy
\eqref{encoded_paulis}. Then, if $U$ is the Clifford operator that
performs the desired deformation on the surface code, we have
\begin{equation}\label{evolucion_paulis_deformacion}
U X_i U\daga = X_{f(s_i)}\simS \prod_j X_j^{a_{ij}},\qquad U Z_i
U\daga  = Z_{f(s_i\prima)}\simS \prod_j Z_j^{d_{ij}},
\end{equation}
where $a_{ij}, d_{ij}\in \Z_2$ and $f(s_i) \simH \sum a_{ij} s_j$,
$f(s_i\prima) \simH \sum d_{ij} s_j\prima$.The equations
\eqref{evolucion_paulis_deformacion} give the evolution of encoded
Pauli operators under $U$ in terms of the continuous map $f$
produced by the deformation related to $U$. It should be noted that
\eqref{evolucion_paulis_deformacion} is very restrictive when
compared to the general \eqref{evolucion_paulis}, so that many
Clifford operations cannot be implemented using deformations. In
particular, Hadamard gates $H$ are out of reach because $H\daga X
H=Z$ and deformations do not mix $X$ and $Z$ operators.

\begin{figure}
\psfrag{s1}{$s_1$}%
\psfrag{s2}{$s_2$}%
\psfrag{s1p}{$s_1\prima$}%
\psfrag{s2p}{$s_2\prima$}%
\psfrag{fs1}{$f(s_1)$}%
\psfrag{fs2p}{$f(s_2\prima)$}%
 \includegraphics[width=13cm]{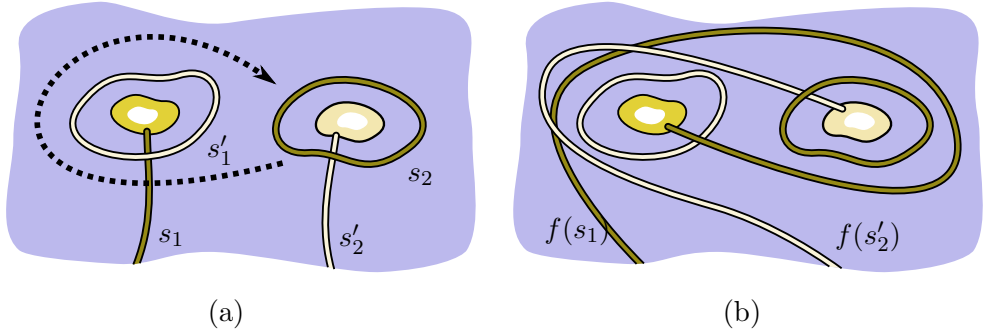}
 \caption{
 A deformation that takes a light hole around a dark one gives rise
 to a CNot gate. (a) A piece of a surface code with a dark and a
 light hole and the nontrivial closed strings of interest.
 (b) After the deformation, two of the strings are mapped to
 different ones.
 }
 \label{figura_CNOT_basica}
\end{figure}

An example is displayed in Fig.~\ref{figura_CNOT_basica}, where the
effect  of moving a light hole around a dark one is analized. There
are two encoded qubits involved in this operation. The deformation
maps the strings $s_1\prima$, $s_2$ to themselves and changes the
strings $s_1$, $s_2\prima$ giving $f(s_1)\simH s_1+s_2$ and
$f(s_2\prima)\simH s_1\prima+s_2\prima$. Taking $X_i:= X_{s_i}$ and
$Z_i:=Z_{s_i\prima}$, $i=1,2$, as the relevant encoded operators,
the result of the deformation is
\begin{equation}\label{evolucion_paulis_CNot}
U X_1 U\daga
 \simS X_1,\qquad U Z_1 U\daga \simS Z_1 Z_2,\qquad U X_2 U\daga \simS X_1 X_2,\qquad U Z_2 U\daga \simS Z_2,
\end{equation}
which corresponds to a CNot gate with the first qubit as source.

\subsection  {Non-smooth deformations}

What happens when a surface code is drastically deformed? Let us
return to the example code of Fig.~\ref{figura_codigo_simple}(a) and
deform it till there exists a nontrivial string of length one, see
Fig.~\ref{figura_codigo_simple}(c). Observe that at this point our
code is absolutely exposed to $X$ errors, but not to $Z$ errors. The
point then is that not only the environment can measure the encoded
$X_1$, but \emph{we also can}. Since $Z$ errors are still unlikely,
the measurement is protected by the code. Although the measurement
seems local, in practice we have to take error correction into
account. For now, we will just concentrate on the effects of
deformations in the absence of errors, and  leave the discussion of
their correction for later. The deformation can be undone, with the
net result that the encoded state has been projected. Thus, we have
succeeded in performing a non-destructive quantum measurement by
code deformation. Visually, the measurement amounts to temporarily
shrink one of the dimensions of the surface. We could have also
employed a similar procedure to measure in the border, as shown in
Fig.~\ref{figura_codigo_simple}(d). In this case one of the dark
borders is contracted.

 The procedures that we have just described involve only smooth deformations, because we did not
change the topology of the surface. However, in
Fig.~\ref{figura_codigo_simple}(c,d)  the removal of a single qubit
would have changed the topology. Because of the topological nature
of the codes and their error correction procedures, it is more
natural to consider non-smooth deformations. For example, let
$\mathcal C$ be the surface code of
Fig.~\ref{figura_codigo_simple}(c) and let $q$ be the qubit that
maintains both pieces of the surface connected. Suppose that we
remove $q$ to obtain a new surface code $\mathcal C\prima$. The dark
square plaquettes that contained $q$ in $\mathcal C$ are triangular
after its removal. We call these new plaquettes $p,p\prima$. Because
we want $C\prima$ to include $q$, we need also an extra stabilizer
to fix it. We choose $X_q$, the $X$ Pauli operator on $q$. Removing
$q$ amounts to cut the surface, and $\mathcal C\prima$ encodes no
qubits, so that we know that the deformation maps some encoded Pauli
operator of the original code to a stabilizer of the new one. In
fact, we can take $U=1$ as the deformation operator. Let
$\ket\psi\in \mathcal C$. Then, if
$X_1\ket\psi=X_q\ket\psi=\ket\psi$, we have $\ket\psi\in\mathcal
C\prima$. On the contrary, if $X_1\ket\psi=X_q\ket\psi=-\ket\psi$
then $\ket\psi\not \in\mathcal C\prima$. In particular, $\ket\psi$
violates the stabilizer conditions for the generators $X_q$, $X_p$
and $X_{p\prima}$. In this sense, in going from $\mathcal C$ to
$\mathcal C\prima$ we are measuring $X_1$.

\begin{figure}
 \includegraphics[width=12cm]{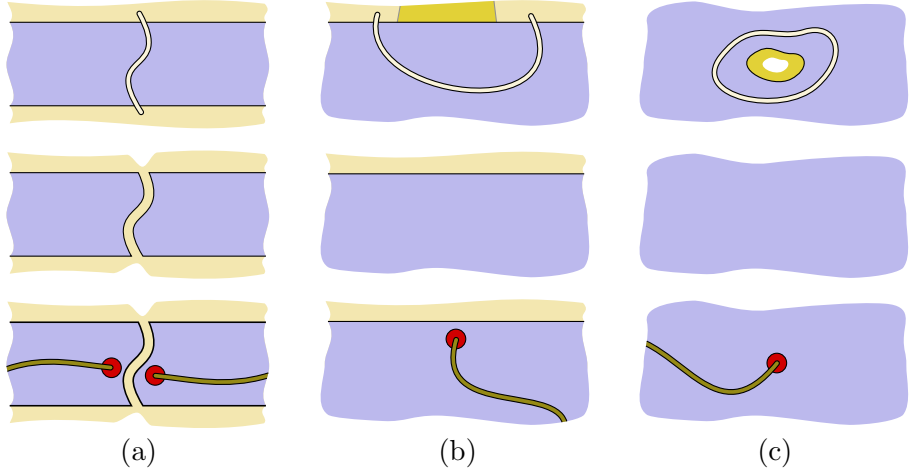}
 \caption
 {
 The effect of non-smooth deformations. The pictures at the top
 part represent the initial situation.
 The $X$ string operator to be measured is displayed as a light string.
 The two pictures below represent the situation afterwards, depending on the measurement
 outcome. The figures in the bottom correspond to the situation in which $X=-1$.
 The stabilizer violations appearing in this case correspond to nonlocal $Z$-strings, as displayed.
 These discontinuous deformations can be read off bottom-up. In that
 case, they represent the introduction of a new encoded qubit,
 together with the initialization $X=1$.
 (a) A cut from border to border. (b) The contraction of a border. (c) The contraction of a hole.
 }
 \label{figura_cut_paste}
\end{figure}

 More generally, cutting a surface along a light string $s$ that connects two different light
borders amounts to measuring $X_s$. For cutting the surface along
$s$ we mean removing all the qubits along it, and we suppose that
the operator $U$ that performs the cut acts locally, in a
neighborhood of the string. To check the previous statement,
consider a light string $s_d$ which is a slight deformation of $s$
lying out of the support of $U$. Then $U X_{s_d}U\daga=X_{s_d}$ and
$X_s\simS X_{s_d}$ . In addition, $X_{s_d}\in \mathcal C\prima$
because $s_d$ is a boundary in the new surface. Moreover, if
$\ket\psi$ is an eigenstate of the stabilizers of $\mathcal
S\prima$, then $X_{s_d}\ket\psi=-\ket\psi$ iff the number of
violations in dark plaquettes in an area with boundary $s_d$ is odd.
Thus, the cut maps the eigenvalue of $X_{s}$ to the parity of the
number of violations that appear at each side of the cut. In the
case of cuts along dark strings $s$, the measured operator is of
course $Z_s$. We can consider also inverse processes, using similar
arguments: If we paste two borders of the lattice together, the new
string operator that runs along the junction is left with definite
eigenvalue $1$. Other non-smooth deformations such as puncturing and
border removal are summarized in Fig.~\ref{figura_cut_paste}.

\subsection {Error correction}

Error correction in surface codes was analyzed in great detail in
\cite{dennis_etal02}. Thus, here we only intend to show how code
deformation can be integrated into the picture given there, which we
summarize now. First, error correction amounts really to keep track
of errors as they show up. To this end, the local generators of the
stabilizer are measured time after time. The results of each round
of measurements gives a time slice, or indeed two, one for
violations of $X$-type generators and one for violations of $Z$-type
generators. For each type, the slices are arranged in a (2+1)-$D$
fashion, where the extra dimension is time. Then, if violations at a
given time are considered as particles, in the (2+1)-$D$ picture we
have their wordlines. In fact, from the measurements we do not get
the actual wordlines. Rather, they have to be inferred, which can be
done correctly with high probability using certain
algorithm\cite{dennis_etal02} as long as errors are below a
threshold. For the procedure to succeed, the actual wordlines and
the inferred ones must be homologically equivalent.

When deformations enter the picture, the error correction procedure
that we have just described is left basically unchanged. When
deformations involve a measurement, that is, the lost of an encoded
qubit, the value of the measurement must be recovered taking into
account the corrections. That is, the relevant stabilizer violations
must be checked after errors have been canceled out. This is
consistent with the fact that only the homology of the wordlines is
relevant.

\subsection {A particular implementation}

There are many ways to encode qubits and to manipulate them through
deformations in a surface code. Here we have chosen an
implementation that is closely connected to the one in
\cite{dennis_etal02}. However, many other implementations could be
given. For example, a disk with a dark external border and $n$ dark
holes encodes $n$ qubits: $X_i$ operators correspond to strings that
enclose the holes and $Z_i$ operators correspond to strings that
connect holes to the external border.

\subsubsection {Encoded qubits and initialization}

\begin{figure}
 \includegraphics[width=14cm]{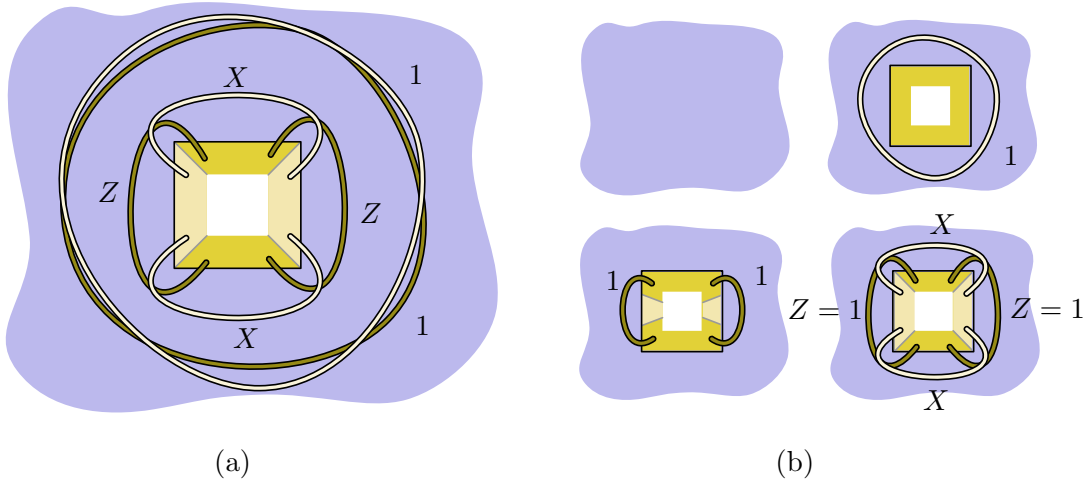}
 \caption
 {
 (a) Each qubit is associated to a hole with
 a dark-light-dark-light border structure.
 Non-trivial operators are labeled with their value.
 (b) The deformation procedure to initialize  a qubit in the state $\ket 1$,
as explained in the text.
 }
 \label{figura_qubits}
\end{figure}

Our starting point is a surface code with an arbitrary shape. As
long as cut and paste operations are performed far enough from the
borders, they are unimportant. The encoded qubits are holes in this
surface, with the particular dark-light-dark-light border structure
depicted in Fig.~\ref{figura_qubits}(a). We impose the condition
that any string operator that surrounds such a hole must have
eigenvalue $1$, something that will be preserved by the gates
proposed below. The encoded $Z$ and $X$ operators can be measured by
shortening a suitable border. As for the initialization procedure
for these encoded qubits in $\ket 1$ ($\ket +$) states, it comprises
two steps. First a dark (light) hole is created by puncturing the
code. Then a pair of light (dark) borders are grown along the border
of the hole. See Fig.~\ref{figura_qubits}(b) for a picture. The
non-smooth operations involved in the procedure are the inverses of
the ones shown in Fig.~\ref{figura_cut_paste}(b,c).

\subsubsection {CNot gate}

\begin{figure}
 \psfrag{X1}{$X_1$}
 \psfrag{X2}{$X_2$}
 \psfrag{Z1}{$Z_1$}
 \psfrag{Z2}{$Z_2$}
 \psfrag{X1X2}{$X_1X_2$}
 \psfrag{Z1Z2}{$Z_1Z_2$}
 \includegraphics[width=14cm]{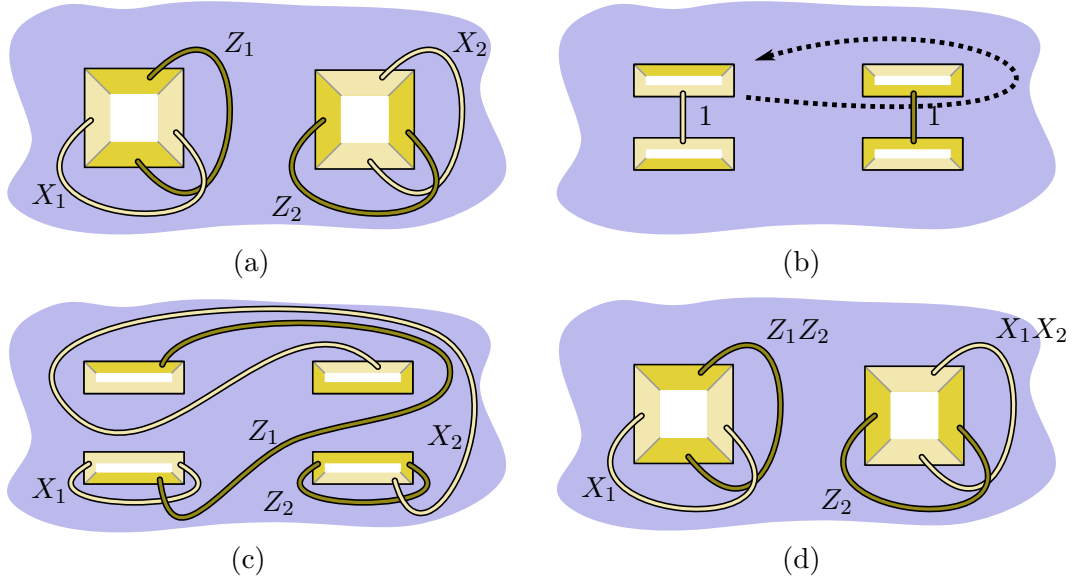}
 \caption{
 The code deformation procedure to obtain a CNot gate. To improve readability, only some of the
 nontrivial strings are shown.
 (a) The qubits prior to the gate:
the left one is the source qubit while the right one is the target.
 (b) A pair of paste operations are performed to obtain two-holed
 qubits.
 The hole movement which is about to be performed is displayed
 dashed.
 (c)
The hole on the top
 of the first qubit winds around one of the
 holes of the second qubit
 and the string operators deform accordingly.
 (d) Finally a pair of cut operations are performed to recover the initial
  configuration. Encoded operators have evolved
  according to the intended CNot.
 }
 \label{figura_cnot}
\end{figure}

The deformation procedure to obtain a CNot gate is explained in
Fig.~\ref{figura_cnot}. It has three steps. First the shape of the
source and target qubits must be altered pasting respectively their
dark and light borders, see Fig.~\ref{figura_cnot}(b). This
operation introduces two new encoded qubits because now new
nontrivial strings exist. Next, one of the holes on the source qubit
winds around one of the holes of the target qubit, as shown in
Figs.~\ref{figura_cnot}(b,c). This step is where the CNot gate
really takes place. Finally, both qubits must recover their original
shape. This step involves cutting along strings, so that the two
qubits that where created in the first step are measured. Comparing
Fig.~\ref{figura_cnot}(a) and Fig.~\ref{figura_cnot}(d) we see that
encoded operators evolve as in \eqref{evolucion_paulis_CNot}, so
that a CNOT is obtained.

\subsubsection {Qubit disconnection and reconnection}

\begin{figure}
 \includegraphics[width=12cm]{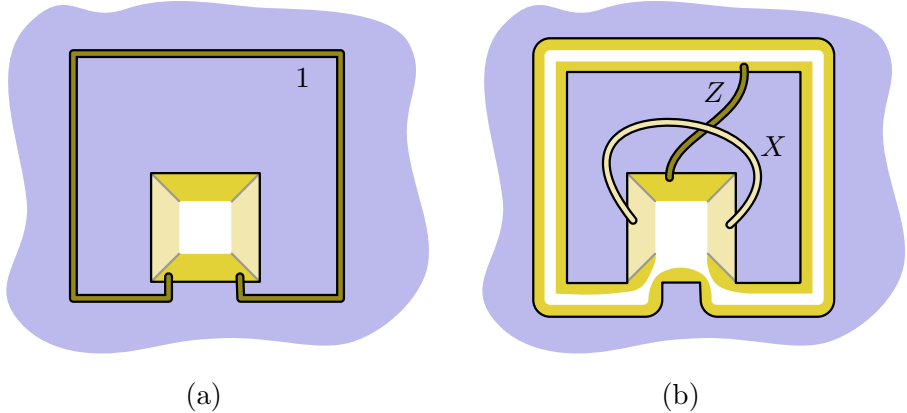}
 \caption{
 The code deformation procedure to disconnect a qubit from the rest of the surface
 code. It can be inverted to reconnect a qubit. (a) The string
 operator along which the cut is done has eigenvalue 1. (b) After the
 cut is done, the resulting isolated qubit lives in the inner
 lattice, which is identical up to deformations to the one in
 Fig.~\ref{figura_codigo_simple}(a).
 }
 \label{figura_desconectar}
\end{figure}

It is useful to disconnect a qubit from the rest of the surface
code. This can be done in several ways, but we choose the one shown
in Fig.~\ref{figura_desconectar}. Observe that the isolated qubit
lives in a lattice equivalent, up to smooth deformations, to the one
in Fig.~\ref{figura_codigo_simple}. These were precisely the qubits
considered in \cite{dennis_etal02}, and thus we can apply all the
single layer processes considered there, such as transversal
initialization of $\ket 0$ and $\ket +$ states, $X$ and $Z$
destructive measurements and encoded Hadamard gates.

\section {Conclusions}

We have introduced code deformation as a tool to perform gates,
initializations and measurements in stabilizer codes. The approach
has been demonstrated in surface codes, where in conjunction with
the techniques discussed in \cite{dennis_etal02} allows to perform
initialization in $\ket +$ or $\ket 1$ states, measurements in $Z$
or $X$ basis, CNot gates and Hadamard gates without exposing any
encoded qubit to errors. It is possible to use these operations to
distill noisy states obtained through progressive 'growth'
\cite{dennis_etal02}. In particular, magic states
\cite{bravyikitaev05} can be distilled in order to perform universal
quantum computation.

\noindent {\em Acknowledgements} We acknowledge financial support
from a PFI fellowship of the EJ-GV (H.B.), DGS grant  under contract
BFM 2003-05316-C02-01 (M.A.MD.), and CAM-UCM grant under ref.
910758.



\begin{thebibliography}{99}

\bibitem{shor95} P. Shor, 1995
Phys. Rev. A {\bf 52}, 2493, 1995.


\bibitem{steane96a}  A.~M. Steane,
Phys. Rev. Lett. {\bf 77} 793, 1996.

\bibitem{gottesman96}
D. Gottesman,
Phys. Rev. A {\bf 54}, 1862 (1996).

\bibitem{calderbank_etal97}
A.~R. Calderbank, E.~M. Rains,  P.~W. Shor,  N.~J.~A. Sloane,
Phys. Rev. Lett. {\bf 78}, 405, (1997).

\bibitem{kitaev97}
A.\,Yu.\,Kitaev,
Annals of Physics \textbf{303} no.~1, 2--30 (2003), \texttt{quant-ph/9707021}.

\bibitem{topologicalclifford}
H. Bombin, M.A. Martin-Delgado;
Phys. Rev. Lett. {\bf 97}, 180501 (2006);
quant-ph/0605138.

\bibitem{tetraUQC}
H. Bombin, M.A. Martin-Delgado;
Phys. Rev. Lett. 98, 160502 (2007); quant-ph/0610024.

\bibitem{dennis_etal02}
E. Dennis, A. Kitaev, A. Landahl, J. Preskill;
J. Math. Phys. {\bf 43}, 4452-4505 (2002).

\bibitem{rauss}
R. Raussendorf, J. Harrington, K. Goyal; New J. Phys. 9 199 (2007);
arXiv:quant-ph/0703143.

\bibitem{AFH08}
R. Alicki, M. Fannes, M. Horodecki;
arXiv:0810.4584

\bibitem{AHHH08}
R. Alicki, M. Horodecki, P. Horodecki, R. Horodecki;
arXiv:0811.0033.


\bibitem{nielsen_chuang}
M. A. Nielsen and I. L. Chuang. "Quantum Computation and Quantum
Information", Cambridge University Press, Cambridge, United Kingdom,
2000.

\bibitem{shor96}
P.W. Shor  in Proc. Symp.  on the Found.  Comp.  Sci. (IEEE press,
Los Alamitos,  California), 56-65, (1996).

\bibitem{knill98}
E. Knill, R. Laflamme, and W. H. Zurek, Philos. Trans. R. Soc.
London, Ser. A 454, 365 (1998); arXiv:quant-ph/9702058.

\bibitem{zeng}
B. Zeng, A. Cross, I. L. Chuang;  arXiv:0706.1382.

\bibitem{bravyi_kitaev_bordes}
S. B. Bravyi, A. Yu. Kitaev; arXiv:quant-ph/9811052.

\bibitem{bmd_homologia}
H. Bombin, M.A. Martin-Delgado, J. Math. Phys. 48, 052105
(2007); arXiv:quant-ph/0605094.

\bibitem{bravyikitaev05}
S. Bravyi, A. Kitaev.
Phys. Rev. A {\bf 71}, 022316 (2005)



\end{thebibliography}
\end{document}